\documentclass[journal]{IEEEtran}
\usepackage{amsmath}
\usepackage{amssymb}
\usepackage{amsfonts}
\usepackage{graphicx}
\usepackage{epsfig}
\usepackage{subfigure}
\usepackage{psfrag}
\usepackage{color}
\usepackage{cases}
\usepackage{cite}

\begin{document}

\title{Channel Diagonalization for Cloud Radio Access}

\author{Liang Liu, \IEEEmembership{Member,~IEEE}, Pratik Patil, and Wei Yu, \IEEEmembership{Fellow,~IEEE}
\thanks{Manuscript received November 24, 2017; revised January 30, 2018;
accepted February 5, 2018. This work was supported by Natural Sciences and Engineering Research Council (NSERC) of Canada. The associate editor coordinating
the review of this paper and approving it for publication was Dr. Tarchi.
(\it{Corresponding author: Liang Liu}.)}
\thanks{
Liang Liu, Pratik Patil, and Wei Yu are with the
Department of Electrical and Computer Engineering, University of Toronto (e-mail:lianguot.liu@utoronto.ca, $\{$ppatil,weiyu$\}$@comm.utoronto.ca).}}

\maketitle

\begin{abstract}
The diagonalization of a conventional multiple-input multiple-output (MIMO)
channel into parallel and independent subchannels via singular value
decomposition (SVD) is a fundamental strategy that allows the MIMO channel
capacity to be achieved using scalar channel codes.
This letter establishes a similar channel diagonalization result for the uplink
and the downlink of a cloud radio access network (C-RAN), in which a central
processor (CP) is connected to a remote radio head (RRH) serving a single user
via rate-limit digital fronthaul carrying the compressed baseband signal.
Specifically, we show that the diagonalization of the MIMO channel between the
RRH and the user via SVD and the subsequent independent and parallel
quantization of scalar signals and channel coding in each of the subchannels is
optimal. This letter establishes this fact using the majorization
theory. Further, an uplink-downlink duality for the multiple-antenna C-RAN is
identified for this single-user case.
\end{abstract}


\setlength{\baselineskip}{1.0\baselineskip}
\newtheorem{definition}{\underline{Definition}}[section]
\newtheorem{fact}{Fact}
\newtheorem{assumption}{Assumption}
\newtheorem{theorem}{Theorem}
\newtheorem{lemma}{\underline{Lemma}}[section]
\newtheorem{corollary}{\underline{Corollary}}[section]
\newtheorem{proposition}{\underline{Proposition}}
\newtheorem{example}{\underline{Example}}[section]
\newtheorem{remark}{\underline{Remark}}[section]
\newtheorem{algorithm}{\underline{Algorithm}}[section]
\newcommand{\mv}[1]{\mbox{\boldmath{$ #1 $}}}

\section{Introduction}\label{sec:Introduction}

A fundamental result in multiple-input multiple-output (MIMO) communication is
that the diagonalization of a Gaussian vector channel
$\mv{y}=\mv{H}\mv{x}+\mv{z}$ via singular value decomposition (SVD),
$\mv{H}=\mv{U}_H\mv{\Lambda}_H\mv{V}_H^H$, leads to an optimal transmission
and reception strategy.  The use of the singular vectors $\mv{V}_H$ as the
precoders, the singular vectors $\mv{U}_H$ as the receive beamformers, and the
use of error-correcting codes designed for single-input single-output (SISO)
additive white Gaussian noise (AWGN) channel over the independent and parallel
subchannels can achieve the capacity of the overall Gaussian vector channel
\cite{raleigh_cioffi}. Moreover, if the uplink and downlink channels
are reciprocals of each other, because the singular values of $\mv{H}$ and
$\mv{H}^H$ are the same, the capacities of the uplink and downlink channels
are also the same.

This letter investigates the uplink and downlink of a cloud radio-access
network (C-RAN) \cite{Simeone16} consisting of a cloud-based central processor (CP), a
remote radio head (RRH), and one user. The RRH is connected to the CP via a finite-capacity
digital link. In the uplink, the RRH compresses the received baseband analog
signal and sends the compressed signal to the CP for demodulation and decoding.
In the downlink, the CP pre-computes the desired analog transmit signal and
sends a compressed version to the RRH for transmission. For such a C-RAN,
% does channel diagonalization via SVD remain optimal?  For such a C-RAN, a reasonable strategy is to
we ask the question of whether the natural strategy of first diagonalizing the MIMO
channel into parallel scalar channels followed by independently compressing the
scalar signal and coding over each of the subchannels is optimal.

This letter answers the question in the affirmative by utilizing
the majorization theory \cite{Olkin79}. The answer is
nontrivial because it involves a joint optimization of transmission and quantization
under finite fronthaul. %and reduces to the well-known channel diagonalization result in \cite{raleigh_cioffi} in the special case of infinite-capacity fronthaul.
As a consequence of the optimality of channel diagonalization, an
uplink-downlink duality is established for the single-RRH single-user C-RAN.

We note that although a main future benefit of C-RAN stems from multiple-RRH
cooperation for interference mitigation, the single-RRH single-user C-RAN
considered here is already of interest, because the processing of baseband
signals in the cloud provides computational offloading benefit
\cite{Dittmann15} even without RRH cooperation. This paper considers C-RAN with
a single user. The single-user C-RAN model considered in this letter can also be thought of as corresponding to
one subcarrier of an orthogonal frequency-division multiple-access
(OFDMA) C-RAN system, in which users are separated via orthogonal subcarriers,
free of co-channel interference \cite{Liu15}.

%Moreover, due to some practical issues such as synchronization, coordinated beamforming among RRHs is challenging for implementation. In this sense, each user is generally served by one adjacent RRH in practice. Hence, our considered setup can be viewed as one subcarrier in the OFDMA-based C-RAN occupied by one user that is served by one RRH.

%\vspace{-5pt}

\section{Uplink and Downlink C-RAN Model}\label{sec:System Model}
\begin{figure*}
\begin{center}
\scalebox{0.55}{\includegraphics*{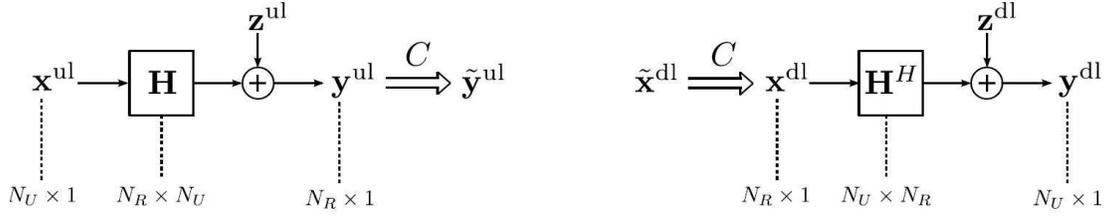}}
\end{center}
\caption{System models of single-user multi-antenna C-RAN in the uplink (left) and
the downlink (right).}\label{fig1}
\end{figure*}

Consider the uplink and downlink C-RAN model consisting of one CP, one RRH and
one user, as shown in Fig. \ref{fig1}. The RRH is equipped with $N_R>1$ antennas; the user is equipped with
$N_U>1$ antennas. In the uplink, the channel from the user to the RRH is
denoted by $\mv{H}\in \mathbb{C}^{N_R\times N_U}$; in the dual downlink, the
channel from  the RRH to the user is given by $\mv{H}^H$ by reciprocity.
Without loss of generality, we assume that the rank of $\mv{H}$ is
$\min(N_R,N_U)$. The
transmit power constraint is denoted by $P$ for both the uplink and the downlink.
It is assumed that the RRH is connected to the CP via a (noiseless) digital
fronthaul link of capacity $C$.%\footnote{\textcolor{red}{Although this letter considers a narrowband channel, the results obtained can be readily extended to broadband based C-RAN
%by viewing each frequency subchannel as one narrowband channel considered
%here and accordingly scaling down the fronthaul capacity by the number of subchannels \cite{Liu15}.}}

%We note that although a main benefit of C-RAN stems from
%multiple-RRH cooperation in which inter-RRH interference can be mitigated,
%the single-RRH single-user C-RAN considered here is already of
%practical interest, since the processing of baseband signals in the cloud
%already brings computational offloading benefit even for this single-RRH case.

This letter considers compression-based strategies for
relaying information between the CP and the RRH via the fronthaul link for both uplink and downlink
\cite{Yuwei13, Simeone13}.
Although from a capacity perspective a RRH capable of decoding and encoding
can potentially outperform compression in our setting,
%considered single-RRH single-user setting,
the compression strategy significantly
simplifies the computational requirements at the RRH. For this reason,
the practical implementations of C-RAN in 5G cellular systems
are likely to involve compression over the fronthaul \cite{Dittmann15}.

\subsection{Uplink C-RAN}

The uplink discrete-time baseband channel is modeled as
\begin{align}
\mv{y}^{\rm ul}=\mv{H}\mv{x}^{\rm ul}+\mv{z}^{\rm ul},
\end{align}
where it is assumed that the user transmits a Gaussian signal with
covariance $\mv{S}^{\rm ul}$, i.e., $\mv{x}^{\rm ul}\sim
\mathcal{CN}(\mv{0},\mv{S}^{\rm ul})$, and $\mv{z}^{\rm ul}\sim
\mathcal{CN}(\mv{0},\sigma^2\mv{I})$ denotes the noise at the RRH.
The total transmit power of the user across all the antennas is
${\rm tr}(\mv{S}^{\rm ul})$.
%\begin{align}\label{eqn:uplink power}
%P^{\rm ul}(\mv{S}^{\rm ul})={\rm tr}(\mv{S}^{\rm ul}).
%\end{align}
The RRH compresses its received signal and sends the compressed signal to the
CP. This quantization process can be modeled as
\begin{align}
\tilde{\mv{y}}^{\rm ul}=\mv{y}^{\rm ul}+\mv{e}^{\rm ul},
\end{align}
where it is assumed that the quantization noise has a Gaussian distribution with
covariance $\mv{Q}^{\rm ul}$, i.e.,
$\mv{e}^{\rm ul}\sim \mathcal{CN}(\mv{0},\mv{Q}^{\rm ul})$.
Based on the rate-distortion theory \cite{Gamal11}, the fronthaul rate needed for
transmitting $\tilde{\mv{y}}^{\rm ul}$ can be expressed as
\begin{align}\label{eqn:uplink fronthaul}
%C^{\rm ul}(\mv{S}^{\rm ul},\mv{Q}^{\rm ul})
I(\mv{y}^{\rm ul};\tilde{\mv{y}}^{\rm ul})
=\log_2\frac{\left|\mv{H}\mv{S}^{\rm ul}\mv{H}^H+\mv{Q}^{\rm ul}+\sigma^2\mv{I}\right|}{\left|\mv{Q}^{\rm ul}\right|}.
\end{align}
The achievable rate for the user, accounting for the quantization noise, can now
be expressed as
\begin{align}\label{eqn:uplink rate}
%R^{\rm ul}(\mv{S}^{\rm ul},\mv{Q}^{\rm ul})=
I(\mv{x}^{\rm ul};\tilde{\mv{y}}^{\rm ul}) =
\log_2\frac{\left|\mv{H}\mv{S}^{\rm ul}\mv{H}^H+\mv{Q}^{\rm ul}+\sigma^2\mv{I}\right|}{\left|\mv{Q}^{\rm ul}+\sigma^2\mv{I}\right|}.
\end{align}

With a transmit power constraint $P$ and fronthaul constraint $C$, the maximum
achievable uplink rate for the user is the optimal value of the following optimization problem \cite{Yuwei13}:
\begin{align*}\mathrm{(P1)}:~\mathop{\mathrm{maximize}}_{\mv{S}^{\rm ul},\mv{Q}^{\rm ul}} & ~~~ \log_2\frac{\left|\mv{H}\mv{S}^{\rm ul}\mv{H}^H+\mv{Q}^{\rm ul}+\sigma^2\mv{I}\right|}{\left|\mv{Q}^{\rm ul}+\sigma^2\mv{I}\right|}   \\
\mathrm {subject \ to} & ~~~ \log_2\frac{\left|\mv{H}\mv{S}^{\rm ul}\mv{H}^H+\mv{Q}^{\rm ul}+\sigma^2\mv{I}\right|}{\left|\mv{Q}^{\rm ul}\right|}\leq C, \\ & ~~~{\rm tr}\left(\mv{S}^{\rm ul}\right)\leq P.\end{align*}

\subsection{Downlink C-RAN}
The downlink discrete-time baseband channel is modeled as
\begin{align}
\mv{y}^{\rm dl}=\mv{H}^H\mv{x}^{\rm dl}+\mv{z}^{\rm dl},
\end{align}where $\mv{x}^{\rm dl}$ denotes the transmit signal of the RRH, and $\mv{z}^{\rm dl}\sim \mathcal{CN}(\mv{0},\sigma^2\mv{I})$ denotes the AWGN at the user.

The transmit signal $\mv{x}^{\rm dl}$ is a compressed version of the beamformed
signal, assumed to be Gaussian distributed as $\tilde{\mv{x}}^{\rm dl}\sim \mathcal{CN}(\mv{0},\tilde{\mv{S}}^{\rm dl})$. The compression process is modeled as
\begin{align}
\mv{x}^{\rm dl}=\tilde{\mv{x}}^{\rm dl}+\mv{e}^{\rm dl},
\end{align}
where the compression noise is also assumed to be Gaussian, %with covariance $\mv{Q}^{\rm dl}$,
i.e., $\mv{e}^{\rm dl}\sim \mathcal{CN}(\mv{0},\mv{Q}^{\rm dl})$.
The total transmit power of the RRH across all the antennas can now be expressed as
${\rm tr}(\tilde{\mv{S}}^{\rm dl}+\mv{Q}^{\rm dl})$.
%\begin{align}\label{eqn:uplink power}
%P^{\rm dl}(\tilde{\mv{S}}^{\rm dl},\mv{Q}^{\rm dl})={\rm tr}(\tilde{\mv{S}}^{\rm dl}+\mv{Q}^{\rm dl}).
%\end{align}
By rate-distortion theory, the fronthaul rate for transmitting $\tilde{\mv{x}}^{\rm dl}$ can be expressed as
\begin{align}\label{eqn:uplink fronthaul}
%C^{\rm dl}(\tilde{\mv{S}}^{\rm dl},\mv{Q}^{\rm dl})=
I(\tilde{\mv{x}}^{\rm dl};\mv{x}^{\rm dl}) =
\log_2\frac{|\tilde{\mv{S}}^{\rm dl}+\mv{Q}^{\rm dl}|}{\left|\mv{Q}^{\rm dl}\right|}.
\end{align}
The achievable rate of the user is
\begin{align}\label{eqn:uplink rate}
%R^{\rm dl}(\tilde{\mv{S}}^{\rm dl},\mv{Q}^{\rm dl})
I(\tilde{\mv{x}}^{\rm dl};\mv{y}^{\rm dl}) =
\log_2\frac{\left|\mv{H}^H\tilde{ \mv{S}}^{\rm dl}\mv{H}+\mv{H}^H \mv{Q}^{\rm dl}\mv{H}+\sigma^2\mv{I}\right|}{\left|\mv{H}^H\mv{Q}^{\rm dl}\mv{H}+\sigma^2\mv{I}\right|}.
\end{align}

With the same transmit power constraint $P$ and fronthaul constraint $C$ as in
the uplink, the maximum achievable downlink rate for the user is the optimal value of the following optimization problem \cite{Simeone13}:
\begin{align*}\mathrm{(P2)}:~\mathop{\mathrm{maximize}}_{\tilde{\mv{S}}^{\rm
dl},\mv{Q}^{\rm dl}} & ~~~ \log_2\frac{\left|\mv{H}^H\tilde{ \mv{S}}^{\rm
dl}\mv{H}+\mv{H}^H \mv{Q}^{\rm
dl}\mv{H}+\sigma^2\mv{I}\right|}{\left|\mv{H}^H\mv{Q}^{\rm
dl}\mv{H}+\sigma^2\mv{I}\right|}   \\
\mathrm {subject \ to} & ~~~ \log_2\frac{|\tilde{\mv{S}}^{\rm dl}+\mv{Q}^{\rm dl}|}{\left|\mv{Q}^{\rm dl}\right|}\leq C, \\ & ~~~{\rm tr}\left(\tilde{\mv{S}}^{\rm dl}+\mv{Q}^{\rm dl}\right)\leq P.\end{align*}

%\begin{figure}
%\begin{center}
%\scalebox{0.4}{\includegraphics*{UplinkDownlinkTransmissionStrategies_MIMO_SingleUserSingleBS.pdf}}
%\end{center}
%\caption{Compress-and-forward strategy for multi-antenna C-RAN.}\label{fig5}
%\end{figure}
%
%Fig. \ref{fig5} illustrates the contrast between the access rate and the fronthaul rate in the uplink and downlink under the compress-and-forward strategy.

\section{Optimality of Channel Diagonalization} \label{sec:Main Result}

This letter shows that the SVD diagonalization of the channel is optimal
for both the uplink and the downlink C-RAN by providing optimal solutions to
problems (P1) and (P2). %We then establish an uplink-downlink duality relationship for the multi-antenna C-RAN.
Define the SVD of $\mv{H}$ as $\mv{H}=\mv{U}_H\mv{\Lambda}_H\mv{V}_H^H$, where $\mv{\Lambda}_H\in \mathbb{C}^{N_R\times N_U}$ is a rectangular diagonal matrix with $D={\rm rank}(\mv{H})=\min(N_R,N_U)$ diagonal elements denoted by $h_1,\cdots,h_D$, with $|h_1|\geq \cdots \geq |h_D|$, and $\mv{U}_H\in \mathbb{C}^{N_R\times N_R}$ and $\mv{V}_H\in \mathbb{C}^{N_U\times N_U}$ are unitary matrices.

\begin{theorem}\label{proposition1}
The optimal solution to problem (P1) has the following form:
\begin{align}
& \mv{S}^{\ast,\rm ul} =\mv{V}_H\mv{\Lambda}_{S^{\rm ul}}\mv{V}_H^H, \label{eqn:optimal uplink transmit}  \\
& \mv{Q}^{\ast,\rm ul} =\mv{U}_H\mv{\Lambda}_{Q^{\rm ul}}\mv{U}_H^H, \label{eqn:optimal uplink compression}
\end{align}where $\mv{\Lambda}_{S^{\rm ul}}={\rm diag}(p^{\rm ul}_1,\cdots,p^{\rm ul}_{N_U})$ and $\mv{\Lambda}_{Q^{\rm ul}}={\rm diag}(q^{\rm ul}_1,\cdots,q^{\rm ul}_{N_R})$ are diagonal matrices with non-negative diagonal elements $p^{\rm ul}_1,\cdots,p^{\rm ul}_{N_U}$ and $q^{\rm ul}_1,\cdots,q^{\rm ul}_{N_R}$.
\end{theorem}

\begin{theorem}\label{proposition2}
The optimal solution to problem (P2) has the following form:
\begin{align}
& \tilde{\mv{S}}^{\ast,\rm dl} =\mv{U}_H\mv{\Lambda}_{\tilde{S}^{\rm dl}}\mv{U}_H^H, \label{eqn:optimal downlink transmit}  \\
& \mv{Q}^{\ast,\rm dl} =\mv{U}_H\mv{\Lambda}_{Q^{\rm dl}}\mv{U}_H^H, \label{eqn:optimal downlink compression}
\end{align}where $\mv{\Lambda}_{\tilde{S}^{\rm dl}}={\rm diag}(\tilde{p}^{\rm dl}_1,\cdots,\tilde{p}^{\rm dl}_{N_R})$ and $\mv{\Lambda}_{Q^{\rm dl}}={\rm diag}(q^{\rm dl}_1,\cdots,q^{\rm dl}_{N_R})$ are diagonal matrices with non-negative diagonal elements $\tilde{p}^{\rm dl}_1,\cdots,\tilde{p}^{\rm dl}_{N_R}$ and $q^{\rm dl}_1,\cdots,q^{\rm dl}_{N_R}$.
\end{theorem}

%\begin{figure}
%\begin{center}
%\scalebox{0.4}{\includegraphics*{MIMO_SingleUserSingleBS_OptimalStructure_Combined.pdf}}
%\end{center}
%\caption{Optimal transmit and compression strategy in the uplink and downlink C-RAN.}\label{fig6}
%\end{figure}
It is well-known that to maximize the capacity of the vector point-to-point
channel, the optimal strategy is to set the precoder and receiver beamformer as
singular vectors in $\mv{V}_H$ and $\mv{U}_H$ so as to diagonalize the channel
$\mv{H}$ \cite{raleigh_cioffi}.  Theorems \ref{proposition1} and \ref{proposition2}
imply that under the assumption of compression-based strategies for C-RAN, the
optimal transmit and receive scheme is also to diagonalize the vector channel
into $D={\rm rank}(\mv{H})$ parallel sub-channels such that a simple quantization
of scalars together with scalar AWGN codes on each subchannel is optimal.
The proofs %of Theorems \ref{proposition1} and \ref{proposition2}
are based on majorization theory \cite{Olkin79} and are deferred to the next section.
Here, $h_d$, $(p^{\rm ul}_d,\tilde{p}^{\rm
dl}_d+q^{\rm dl}_d)$, and $(q^{\rm ul}_d,q^{\rm
dl}_d)$ are the channel strength, transmit power, and quantization
noise level of the uplink/downlink subchannel $d$. %respectively.

The optimality of channel diagonalization allows us to establish
an uplink-downlink duality relationship for the multi-antenna single-user
single-RRH C-RAN.  Based on Theorems \ref{proposition1} and \ref{proposition2},
the maximizations of the uplink and downlink rates for C-RAN, i.e., (P1) and (P2),
reduce to the following scalar optimization problems. For the uplink, we have:
\begin{align}~\mathop{\mathrm{maximize}}_{\{p^{\rm ul}_d,q^{\rm ul}_d\}} & ~~~
\sum\limits_{d=1}^D\log_2\left(1+\frac{|h_d|^2p^{\rm
ul}_d}{q^{\rm ul}_d+\sigma^2}\right) \label{eqn:problem P1'}  \\
\mathrm {subject \ to} & ~~~ \sum\limits_{d=1}^D\log_2\left(\frac{|h_d|^2p^{\rm ul}_d+q^{\rm ul}_d+\sigma^2}{q^{\rm ul}_d}\right)\leq C, \nonumber \\ & ~~~\sum\limits_{d=1}^Dp^{\rm ul}_d\leq P, \nonumber \end{align}
and for the downlink, we have:
\begin{align}~\mathop{\mathrm{maximize}}_{\{\tilde{p}^{\rm dl}_d,q^{\rm dl}_d\}} & ~~~ \sum\limits_{d=1}^D\log_2\left(1+\frac{|h_d|^2\tilde{p}^{\rm dl}_d}{|h_d|^2q^{\rm dl}_d+\sigma^2}\right) \label{eqn:problem P2'} \\
\mathrm {subject \ to} & ~~~ \sum\limits_{d=1}^D\log_2\left(\frac{\tilde{p}^{\rm dl}_d+q^{\rm dl}_d}{q^{\rm dl}_d}\right)\leq C, \nonumber \\ & ~~~\sum\limits_{d=1}^D(\tilde{p}^{\rm dl}_d+q^{\rm dl}_d)\leq P. \nonumber \end{align}
%Unfortunately, both problems (\ref{eqn:problem P1'}) and (\ref{eqn:problem P2'}) are non-convex problems. Suboptimal solutions can be obtained by alternating optimization techniques since it can be shown that if the power or quantization noise level is fixed, the above problems are convex over the other set of variables.
Although the above problems (\ref{eqn:problem P1'}) and (\ref{eqn:problem P2'})
are not convex,
%non-convex problems. Suboptimal solutions can be obtained by alternating optimization techniques since it can be shown that if the power or quantization noise level is fixed, the above problems are convex over the other set of variables.
%Although we cannot solve problems (\ref{eqn:problem P1'}) and (\ref{eqn:problem P2'}) globally, there
there is an interesting duality between the two. For the SISO C-RAN, the duality has been shown in
\cite[Theorem 1]{Liu16}.
%where the RRH and the user are equipped with one antenna, there is a duality between the uplink and the downlink \cite[Theorem 1]{Liu16}.
Specifically, consider the case of $D=1$,
%in (\ref{eqn:problem P1'}),
the maximum uplink achievable rate is %(omitting $d$):
\begin{align}
\log_2\left(1+\frac{|h|^2p^{\rm ul}}{q^{\rm ul}+\sigma^2}\right) =
\log_2\left(\frac{|h|^2 P +\sigma^2} {\sigma^2+2^{-C}|h|^2 P}\right),
\end{align}
where we have used the relationships $p^{\rm ul}=P$ and
\begin{align}
\log_2\left(\frac{|h|^2p^{\rm ul}+q^{\rm ul}+\sigma^2}{q^{\rm ul}}\right) = C.
\end{align}
Likewise in the downlink $D=1$ case, utilizing the relationships
$\tilde{p}^{\rm dl}+q^{\rm dl} = P$ and
\begin{align}
\log_2\left(\frac{\tilde{p}^{\rm dl}+q^{\rm dl}}{q^{\rm dl}}\right)= C,
\end{align}
it can be shown that the maximum downlink achievable rate is exactly the same as the uplink:
\begin{align}
\log_2\left(1+\frac{|h|^2\tilde{p}^{\rm
dl}}{|h|^2q^{\rm dl}+\sigma^2}\right) =
\log_2\left(\frac{|h|^2 P +\sigma^2} {\sigma^2+2^{-C}|h|^2 P}\right).
\end{align}
By applying this result to each subchannel $d$, it can be shown that the
optimal value of the problem (\ref{eqn:problem P1'}) must be the same as that
of the problem (\ref{eqn:problem P2'}).
%As a result, the optimal value of the problem (P1) is the same as that of the problem (P2), thus establishing an uplink-downlink duality. %Note that the rate region for the single-user setup is a continuous segment for both the uplink and downlink.  Since the extreams of these two segments are the same, we have the following theorem.
\begin{theorem}\label{theorem2}
Consider a C-RAN model implementing compression-based strategies in both the uplink and the downlink, where both the user and RRH are equipped with multiple antennas. Any rate that is achievable in the uplink is achievable with the same sum-power and fronthaul capacity constraint in the downlink, and vice versa.
\end{theorem}

The above result can be thought of as a generalization of celebrated
uplink-downlink duality (e.g.\ \cite{Goldsmith03}) to the C-RAN, but restricted
to the single-RRH single-user case.
%As discussed in \cite{Liu16},
%In general the resource allocation problems in the uplink are easier to solve than those in the downlink. As a result,

% Theorem \ref{theorem2} can be used to solve resource allocation problems in the downlink C-RAN based on their counterpart in the uplink, which is typically easier to solve.

\section{Proofs via Logarithmic Majorization}

\subsection{Proof of Theorem \ref{proposition1}}\label{appendix3}
First by \cite[Theorem 2]{Yu15}, we can recast problem (P1) as
\begin{align}\mathop{\mathrm{maximize}}_{R,{\mathbf S}^{\rm ul},{\mathbf Q}^{\rm ul}} & ~~~ R  \label{eqn:problem 1} \\
\mathrm {subject \ to} & ~~~ R\leq \log_2\left|\mv{I}+\mv{H}\mv{S}^{\rm ul}\mv{H}^H(\mv{Q}^{\rm ul}+\sigma^2\mv{I})^{-1}\right|, \nonumber \\ & ~~~R+\log_2\frac{\left|\mv{Q}^{\rm ul}+\sigma^2\mv{I}\right|}{\left|\mv{Q}^{\rm ul}\right|}\leq C, \nonumber \\ & ~~~{\rm tr}\left(\mv{S}^{\rm ul}\right)\leq P.\nonumber \end{align}Define the SVD of $\mv{\Phi}=\mv{H}\mv{S}^{\rm ul}\mv{H}^H$ and $\mv{Q}^{\rm ul}$ as $\mv{\Phi}=\mv{U}_{\Phi}\mv{\Lambda}_{\Phi}\mv{U}_{\Phi}^H$ and $\mv{Q}^{\rm ul}=\mv{U}_{Q^{\rm ul}}\mv{\Lambda}_{Q^{\rm ul}}\mv{U}_{Q^{\rm ul}}^H$, respectively, for some $\mv{\Lambda}_{\Phi}={\rm diag}(\lambda_{\Phi,1},\cdots,\lambda_{\Phi,N_R})$ with $\lambda_{\Phi,1}\geq \cdots \geq \lambda_{\Phi,N_R}$, and $\mv{\Lambda}_{Q^{\rm ul}}={\rm diag}(\lambda_{U^{\rm ul},1},\cdots,\lambda_{U^{\rm ul},N_R})$ with $\lambda_{U^{\rm ul},1}\leq \cdots \leq \lambda_{U^{\rm ul},N_R}$.
Note that the optimization variables of problem (\ref{eqn:problem 1}) are changed to $R$, $\mv{\Phi}$ (instead of $\mv{S}^{\rm ul}$), and $\mv{Q}^{\rm ul}$.

%Then, problem (\ref{eqn:problem 1}) can be transformed into the following problem.
%\begin{align}\mathop{\mathtt{Maximize}}_{R,\mv{U}_{\Phi},\mv{\Lambda}_{\Phi},\mv{U}_{Q^{\rm ul}},\mv{\Lambda}_{Q^{\rm ul}}} & ~~~ R  \label{eqn:problem 2} \\
%\mathtt {Subject \ to} & ~~~ R\leq \log_2\left|\mv{I}+\mv{U}_{\Phi}\mv{\Lambda}_{\Phi}\mv{U}_{\Phi}^H%\mv{U}_{Q^{\rm ul}}(\mv{\Lambda}_{Q^{\rm ul}}+\sigma^2\mv{I})^{-1}\mv{U}_{Q^{\rm ul}}^H\right|, \nonumber \\ & ~~~R+\log_2\frac{\left|\mv{\Lambda}_{Q^{\rm ul}}+\sigma^2\mv{I}\right|}{|\mv{\Lambda}_{Q^{\rm ul}}|}\leq C, \nonumber \\ & ~~~{\rm tr}(\mv{U}_{\Phi}\mv{\Lambda}_{\Phi}\mv{U}_{\Phi}^H\mv{U}_H(\mv{\Lambda}_H\mv{\Lambda}_H^H)^{-1}\mv{U}_H^H)\leq P.\nonumber \end{align}

Consider the first constraint in problem (\ref{eqn:problem 1}). Given any matrix $\mv{A}$, define $\mv{\gamma}^\downarrow(\mv{A})$ and $\mv{\gamma}^\uparrow(\mv{A})$ as the vectors consisting of all the eigenvalues of $\mv{A}$ in decreasing and increasing order, respectively. According to \cite[9.H.1.d]{Olkin79}, $\mv{\gamma}(\mv{\Phi}(\mv{Q}^{\rm ul}+\sigma^2\mv{I})^{-1})$ is logarithmically majorized by $\mv{\gamma}^\downarrow(\mv{\Phi})\circ\mv{\gamma}^\downarrow((\mv{Q}^{\rm ul}+\sigma^2\mv{I})^{-1})$ (see \cite[Definition 1.4]{Guan07}), where $\circ$ represents the elementwise product. Moreover, it can be shown that $f(\mv{x})=\sum_{n=1}^N\log_2(\sigma^2+x_n)$ is a Schur-geometrically-convex function of $\mv{x}=[x_1,\cdots,x_N]^T$ (see \cite[Definition 1.5]{Guan07}). Since the diagonal elements of $\mv{\Lambda}_{\Phi}$, $(\mv{\Lambda}_{Q^{\rm ul}}+\sigma^2\mv{I})^{-1}$ are all arranged in deceasing order, we thus have
\begin{align}\label{eqn:1}
& \log_2\left|\mv{I}+\mv{H}\mv{S}^{\rm ul}\mv{H}^H(\mv{Q}^{\rm ul}+\sigma^2\mv{I})^{-1}\right| \nonumber \\ = & \log_2\left|\mv{I}+\mv{\Phi}(\mv{Q}^{\rm ul}+\sigma^2\mv{I})^{-1}\right| \nonumber \\ = & \log_2|\sigma^2\mv{I}+{\rm diag}(\mv{\gamma}(\mv{\Phi}(\mv{Q}^{\rm ul}+\sigma^2\mv{I})^{-1}))| \nonumber \\ \leq & \log_2|\sigma^2\mv{I}+{\rm diag}(\mv{\gamma}^\downarrow(\mv{\Phi})\circ\mv{\gamma}^\downarrow((\mv{Q}^{\rm ul}+\sigma^2\mv{I})^{-1}))| \nonumber \\ = & \log_2|\sigma^2\mv{I}+\mv{\Lambda}_{\Phi}(\mv{\Lambda}_{Q^{\rm ul}}+\sigma^2\mv{I})^{-1}|,
\end{align}where the equality holds if and only if $\mv{U}_{Q^{\rm ul}}=\mv{U}_{\Phi}$.

Next, consider the second constraint in problem (\ref{eqn:problem 1}). It can
be observed that it does not depend on $\mv{U}_{\Phi}$ and $\mv{U}_{Q^{\rm ul}}$.

Finally, consider the third constraint in problem (\ref{eqn:problem 1}).
Define the truncated SVD
of $\mv{H}$ as $\mv{H}=\bar{\mv{U}}_H\bar{\mv{\Lambda}}_H\bar{\mv{V}}_H^H$,
where $\bar{\mv{U}}_H$ and $\bar{\mv{V}}_H$ consist of the first
$D=\min(N_R,N_U)$ columns of $\mv{U}_H$ and $\mv{V}_H$ with
$\bar{\mv{U}}_H^H\bar{\mv{U}}_H=\bar{\mv{V}}_H^H\bar{\mv{V}}_H=\mv{I}$, and
$\bar{\mv{\Lambda}}_H\in \mathbb{C}^{D\times D}$ is the block-submatrix of
$\mv{\Lambda}_H$ with indices taken from $1$ to $D$.  Since
$\mv{\Phi}=\mv{H}\mv{S}^{\rm ul}\mv{H}^H$, we have $\bar{\mv{V}}_H^H\mv{S}^{\rm
ul}\bar{\mv{V}}_H=\bar{\mv{\Lambda}}_H^{-1}\bar{\mv{U}}_H^H\mv{\Phi}\bar{\mv{U}}_H\bar{\mv{\Lambda}}_H^{-1}$.
It then follows that
\begin{align}\label{eqn:5}
{\rm tr}\left(\mv{S}^{\rm ul}\right) & \overset{(a)}{\geq}{\rm tr}(\mv{S}^{\rm ul}\bar{\mv{V}}_H\bar{\mv{V}}_H^H) \nonumber \\ & ={\rm tr}\left(\mv{\Phi}\bar{\mv{U}}_H\bar{\mv{\Lambda}}_H^{-2}\bar{\mv{U}}_H^H\right)\nonumber \\ & ={\rm tr}\left(\mv{\Phi}\mv{U}_H(\mv{\Lambda}_H\mv{\Lambda}_H^H)^{-1}\mv{U}_H^H\right),
\end{align}where $(a)$ is from \cite[Appendix A]{Goldsmith03}. Since the diagonal elements of $(\mv{\Lambda}_{H}\mv{\Lambda}_H^H)^{-1}$ are arranged in increasing order, according to \cite[9.H.1.h]{Olkin79}, we have
\begin{align}\label{eqn:2}
{\rm tr}\left(\mv{\Phi}\bar{\mv{U}}_H(\mv{\Lambda}_H\mv{\Lambda}_H^H)^{-1}\bar{\mv{U}}_H^H\right) & \geq {\rm tr}\left(\mv{\Lambda}_{\Phi} (\mv{\Lambda}_H\mv{\Lambda}_H^H)^{-1}\right) \nonumber \\ & = \sum\limits_{d=1}^{N_R}\frac{\lambda_{\Phi,d}}{|h_d|^2}.
\end{align}
Note that if $h_d=0$, for $d=D+1,\cdots,N_R$, we have
$\lambda_{\Phi,d}=0$ and $\frac{\lambda_{\Phi,d}}{|h_d|^2}$ is
understood to be zero.
Combining (\ref{eqn:5}) and (\ref{eqn:2}), it follows that the transmit power
is lower-bounded by $\sum_{d=1}^{N_R}\lambda_{\Phi,d}/|h_d|^2$.
Observe that
if $\mv{U}_{\Phi}=\mv{U}_H$, or equivalently, the SVD of $\mv{S}^{\rm}$ is
in the form of (\ref{eqn:optimal uplink transmit}),  the above lower bound on
transmit power is achieved, i.e., ${\rm tr}(\mv{S}^{\rm ul})=\sum_{d=1}^{N_R}\lambda_{\Phi,d}/|h_d|^2$.

Further,
regardless of the choice of $\mv{U}_{\Phi}$, the upper bound of
$\log_2\left|\mv{I}+\mv{H}\mv{S}^{\rm ul}\mv{H}^H(\mv{Q}^{\rm
ul}+\sigma^2\mv{I})^{-1}\right|$ given in (\ref{eqn:1}) can be achieved as long
as the optimal $\mv{U}_{Q^{\rm ul}}$ is set to be $\mv{U}_{\Phi}$. As a
consequence,
the optimal solution is to set $\mv{U}_{\Phi}$ as $\mv{U}_H$, i.e.,
(\ref{eqn:optimal uplink transmit}), since it makes the transmit power the lowest,
and to set
$\mv{U}_{Q^{\rm ul}}=\mv{U}_{\Phi}=\mv{U}_H$ to achieve the maximum value of
$\log_2\left|\mv{I}+\mv{H}\mv{S}^{\rm ul}\mv{H}^H(\mv{Q}^{\rm
ul}+\sigma^2\mv{I})^{-1}\right|$. This makes the feasible set of $R$,
$\mv{\Lambda}_{S^{\rm ul}}$, and $\mv{\Lambda}_{Q^{\rm ul}}$ in problem
(\ref{eqn:problem 1}) the largest, which results in the best objective value.
Theorem \ref{proposition1} is thus proved.

\subsection{Proof of Theorem \ref{proposition2}}\label{appendix4}
For convenience, define $\mv{S}^{\rm dl}=\tilde{\mv{S}}^{\rm dl}+\mv{Q}^{\rm dl}$ as the covariance matrix of the transmit signal $\mv{x}^{\rm dl}$. Problem (P2) reduces to:
\begin{align}\mathop{\mathrm{maximize}}_{{\mathbf S}^{\rm dl},{\mathbf Q}^{\rm dl}} & ~~~ \log_2\frac{\left|\mv{H}^H \mv{S}^{\rm dl}\mv{H}+\sigma^2\mv{I}\right|}{\left|\mv{H}^H\mv{Q}^{\rm dl}\mv{H}+\sigma^2\mv{I}\right|}   \label{eqn:problem 5} \\
\mathrm {subject \ to} & ~~~\log_2\frac{\left|\mv{S}^{\rm dl}\right|}{\left|\mv{Q}^{\rm dl}\right|}\leq C, \nonumber \\ & ~~~{\rm tr}\left(\mv{S}^{\rm dl}\right)\leq P. \nonumber \end{align}

First, given any $\mv{Q}^{\rm dl}=\bar{\mv{Q}}^{\rm dl}$, we optimize over $\mv{S}^{\rm dl}$. Define the SVD of $\mv{S}^{\rm dl}=\mv{U}_{S^{\rm dl}}\mv{\Lambda}_{S^{\rm dl}}\mv{U}_{S^{\rm dl}}^H$, where $\mv{\Lambda}_{S^{\rm dl}}={\rm diag}(p^{\rm dl}_1,\cdots,p^{\rm dl}_{N_R})$ with $p^{\rm dl}_1\geq \cdots \geq p^{\rm dl}_{N_R}$. Then, problem (\ref{eqn:problem 5}) reduces to the following problem:
\begin{align}\mathop{\mathrm{maximize}}_{{\mathbf U}_{S^{\rm dl}},{\mathbf
\Lambda}_{S^{\rm dl}}} & ~~~ \log_2|\mv{U}_{S^{\rm dl}}\mv{\Lambda}_{S^{\rm
dl}}\mv{U}_{S^{\rm
dl}}^H\mv{U}_H\mv{\Lambda}_H\mv{\Lambda}_H^H\mv{U}_H^H+\sigma^2\mv{I}| \nonumber \\
\mathrm {subject \ to} & ~~~ \log_2|\mv{\Lambda}_{S^{\rm dl}}|\leq C+\log_2|\bar{\mv{Q}}^{\rm dl}|, \nonumber \\ & ~~~{\rm tr}(\mv{\Lambda}_{S^{\rm dl}})\leq P.
\label{eqn:problem 3}
\end{align}

Note that the diagonal elements of $\mv{\Lambda}_{S^{\rm dl}}$ and $\mv{\Lambda}_H\mv{\Lambda}_H^H$ are arranged in decreasing order. Similar to (\ref{eqn:1}), we have
\begin{multline}
\log_2|\mv{U}_{S^{\rm dl}}\mv{\Lambda}_{S^{\rm dl}}\mv{U}_{S^{\rm dl}}^H\mv{U}_H\mv{\Lambda}_H\mv{\Lambda}_H^H\mv{U}_H^H+\sigma^2\mv{I}| \\ \leq \log_2|\mv{\Lambda}_{S^{\rm dl}}\mv{\Lambda}_H\mv{\Lambda}_H^H+\sigma^2\mv{I}|,
\label{eq:opt_S}
\end{multline}
where the equality holds if and only if $\mv{U}_{S^{\rm dl}}=\mv{U}_H$.

Next, given any $\mv{S}^{\rm dl}=\bar{\mv{S}}^{\rm dl}$, we optimize $\mv{Q}^{\rm dl}$. Define $\mv{Q}^{\rm dl}=\mv{U}_{Q^{\rm dl}}\mv{\Lambda}_{Q^{\rm dl}}\mv{U}_{Q^{\rm dl}}^H$, where $\mv{\Lambda}_{Q^{\rm dl}}={\rm diag}(q^{\rm dl}_1,\cdots,q^{\rm dl}_{N_R})$ with $q^{\rm dl}_1\leq \cdots \leq q^{\rm dl}_{N_R}$.
Then, problem (\ref{eqn:problem 5}) reduces to the following problem:
\begin{align}\mathop{\mathrm{minimize}}_{{\mathbf U}_{Q^{\rm dl}},{\mathbf \Lambda}_{Q^{\rm dl}}} & ~~~
\log_2|\mv{U}_H\mv{\Lambda}_H\mv{\Lambda}_H^H\mv{U}_H^H\mv{U}_{Q^{\rm
dl}}\mv{\Lambda}_{Q^{\rm dl}}\mv{U}_{Q^{\rm dl}}^H+\sigma^2\mv{I}|  \nonumber \\
\mathrm {subject \ to} & ~~~ \log_2|\mv{\Lambda}_{Q^{\rm dl}}|\geq \log_2|\bar{\mv{S}}^{\rm dl}|-C.  \label{eqn:problem 4}
\end{align}
According to \cite[III.6.14]{Bhatia97}, $\mv{\gamma}^\downarrow(\mv{H}\mv{H}^H)\circ\mv{\gamma}^\uparrow(\mv{Q}^{\rm dl})$ is logarithmically majorized by $\mv{\gamma}(\mv{H}\mv{H}^H\mv{Q}^{\rm dl})$.
Then, we have
\begin{multline}
\log_2|\mv{U}_H\mv{\Lambda}_H\mv{\Lambda}_H^H\mv{U}_H^H\mv{U}_{S^{\rm dl}}\mv{\Lambda}_{Q^{\rm dl}}\mv{U}_{S^{\rm dl}}^H+\sigma^2\mv{I}| \\ \geq \log_2|\mv{\Lambda}_H\mv{\Lambda}_H^H\mv{\Lambda}_{Q^{\rm dl}}+\sigma^2\mv{I}|,
\label{eq:opt_Q}
\end{multline}
where the equality holds if and only if $\mv{U}_{Q^{\rm dl}}=\mv{U}_H$.

Combining (\ref{eq:opt_S}) and (\ref{eq:opt_Q}), Theorem \ref{proposition2} is thus proved.

\section{Conclusion}
\label{sec:Conclusion}
This letter shows that for the single-RRH single-user multi-antenna C-RAN,
where the compression-based strategies are employed to relay information
between the RRH and the CP, for
both the uplink and the downlink, the optimal strategy is to perform
the channel SVD-based linear precoding and receive beamforming to diagonalize
the MIMO channel into parallel SISO subchannels then perform compression and
channel coding on each subchannel independently. %This is a generalization of the celebrated capacity-achieving result for the MIMO point-to-point channel.
This result leads to an uplink-downlink duality for the single-user C-RAN.

\end{document}